%% file: report.tex
\documentclass{tech-report}
\usepackage{blindtext}
\usepackage{graphicx}
\usepackage{fancyhdr,graphicx,amsmath,amssymb}
\usepackage[ruled,vlined]{algorithm2e}
\usepackage{afterpage}
\usepackage{pdfpages}

\newcommand\blankpage{%
    \null
    \thispagestyle{empty}%
    \addtocounter{page}{-1}%
    \newpage}
    
\include{pythonlisting}

\title{Private Queries on Public Certificate Transparency Data}
\author{by Vy-An Phan}

\bibliography{bibliography}

\begin{document}

\includepdf[page={1,2,3}]{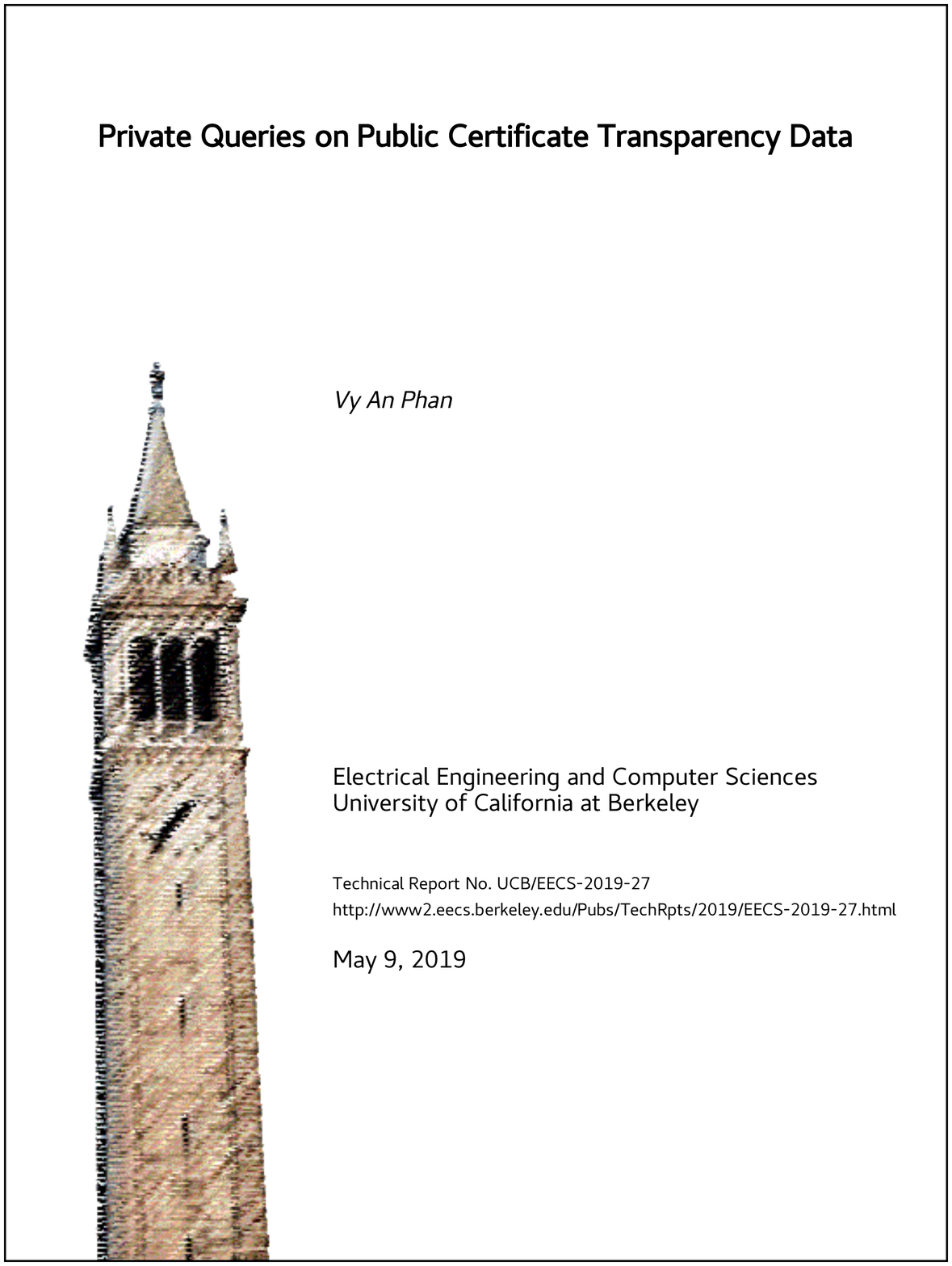}
\afterpage{\blankpage}
\include{src/abstract}
\tableofcontents

\include{src/introduction}

\include{src/relwork}
\include{src/design}
\include{src/implementation}
\include{src/eval}
\include{src/conclusion}
\include{src/acknowledgements}

\bookmarksetup{startatroot}
\printbibliography[heading=bibintoc]

\end{document}

%% file: src/abstract.tex
\chapter*{Abstract}
\label{sec:abstract}

Despite increasing advancements in today's information exchange infrastructure, the preservation of user data and privacy still remains a problem. Both insecure baselines and secure solutions leak user data. For example, Certificate Transparency (CT) promises significant security improvements to existing Public Key Infrastructure solutions that up-to-now have solely relied on the Certificate Authority hierarchy. CT provides a robust auditing layer and transparency solution to quickly detect such compromises, but introduces the requirement that client browsers interact with third-party servers when validating a site certificate.

In the existing CT system, these requests leak information about each user's browsing habits to the hosting server. It is not a stretch to think that this valuable data could be collected and exploited, as corporations and governments have plenty of financial and political incentive to do so. In this project, we seek to address this problem by using an oblivious file sharing system with strong anonymity properties, to provide a more scalable, performant solution to privacy-preserving queries.

%% file: src/introduction.tex
\chapter{Introduction}
\label{sec:intro}

Modern secure communication between users and websites requires on the exchange of public keys. The Public Key Infrastructure (PKI) is a core underpinning technology of the modern internet, enabling secure traffic (e.g. via TLS) between clients and servers.

Traditionally, users depended on central certificate authorities (CAs) for issuing, moderating, and verifying certificates. This model of putting all trust into a single point of failure has proven to be far from perfect. There are documented cases of CAs incorrectly issuing certificates~\cite{google-mitm,misissuance-wild}, being attacked by malicious third parties~\cite{comodo-incident}, and even engaging in malicious behavior themselves~\cite{google-mistrust}. Even worse, failure in the CA system is an "or" condition, not an "and" condition. There are dozens of CAs in existence, and it only takes one of them to be compromised in order for all the sites dependent on it to be compromised, even if said websites had their public keys stored with other CAs. These incidents can expose large numbers of users to active man-in-the-middle (MITM) attacks. 

In response to these growing concerns, Certificate Transparency (CT) is a new protocol that addresses these issues by enabling users, CAs, and domain owners to audit certificate changes and identify and monitor potential modifications directly relevant to them~\cite{ct}. Recently, CT has seen uptake by both web browsers and CAs, with Google Chrome requiring all trusted CAs to partake in the protocol since 2017~\cite{chrome-requirement}. All changes are logged via hashed chains of Merkle tree roots, preventing editing of prior history and allowing easy auditing of future changes. 

\begin{figure}
  \includegraphics[width=\linewidth]{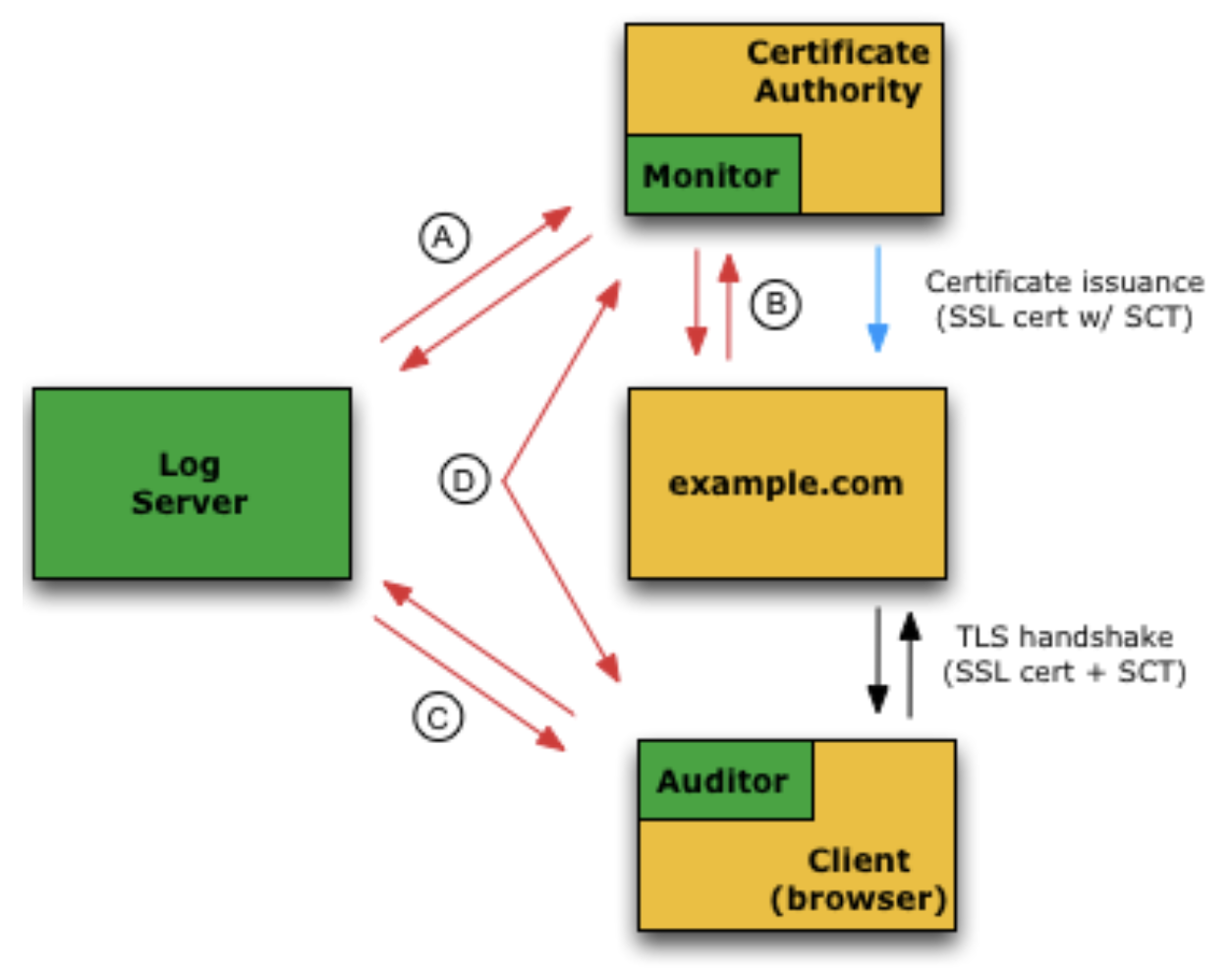}
  \caption{A simplified architecture diagram for Certificate Transparency. Assuming an honest-but-curious log server, request C is problematic as it reveals to the log server what sites the client is visiting.}
  \label{fig:ct_arch}
\end{figure}

However, CT is not without its flaws. In particular, it represents an additional communication channel --- as part of normal operation, users query servers hosting CT log for certificates matching domains with which they wish to communicate as shown in Figure~\ref{fig:ct_arch}. This querying is a central part of the CT protocol, but has the side effect that a malicious log server can glean private information about the user from their queries, namely, correlating the IP address of the request with the domains visited over time.

Such tracking could easily be performed without consent. Particularly worrisome is that these log servers have financial incentive to do exactly that: transparently collect user browsing data and sell it to industry players in advertisement and tracking. This example is just one of many instances that illustrate the privacy concerns that arise from requiring users to contact a third party (CT logs) to communicate with any desired domain. In general, users should be able to benefit from CT without sacrificing their privacy.

Previous efforts to maintain privacy in Certificate Transparency and other data retrieval systems are discussed in Section~\ref{sec:relwork}. In this project, we seek to extend the CT protocol to preserve users' privacy from the log servers that they query. 

We outline our definition of privacy with subsequent design goals in this setting in Section~\ref{sec:design}. Our solution should be usable and scalable, in order to efficiently service hundreds of millions of users, without sacrificing user privacy by revealing their queries to the server.

In Section~\ref{sec:implementation} we describe our solution. We use an ORAM file sharing system~\cite{oram-filesharing} to achieve the base security properties, and add system-level performance improvements to make it more usable. We do this by precomputing encryption exponentiations and garbled circuits, pipelining phases of the transfer protocol, and batching requests. 

Despite implementing all of our improvements, we discovered that we did not meet our scalability goals. Even so, we saw better performance that points to a possibility of meeting those goals in the future. This will be further discussed in Section~\ref{sec:eval}.

%% file: src/relwork.tex
\chapter{Related Work}
\label{sec:relwork}

\section{Privacy in Certificate Transparency}

Since Certificate Transparency's inception, there have been significant extensions to increase its functionality and security. For example, CT has been extended to support certificate revocation and subsequent applications including end-to-end encrypted mail~\cite{enhanced-ct}.

Additionally, the security guarantees of CT have been demonstrated via cryptographic properties~\cite{secure-logging} and measurement studies have been conducted to evaluate how it is used in practice and the differences between public CT logs~\cite{ct-landscape}.

However, evaluation and extension of the privacy properties of CT have not been as thoroughly scrutinized. Eskandarian et al.\ identified and proposed a solution to privacy issues related to browsers auditing a CT log and extended CT to support non-public subdomains~\cite{ct-with-privacy}. To the best of our knowledge, however, there does not exist an extension to CT that preserves users' privacy when interacting with CT logs.

\section{Oblivious RAM}

An adversary with access to a machine's data access patterns can gain nontrivial information about a particular process, even if the data itself is encrypted. Oblivious RAM (ORAM)~\cite{oram} schemes prevent this by randomizing data access patterns in such a way that observing them reveals no information about the data being accessed. This is especially useful in the case where a client or user wishes to access data on a server or machine without revealing what that data is. 

Primitive approaches to ORAM are not very efficient. Path ORAM~\cite{path-oram} is an improvement on traditional ORAM by storing data blocks in a tree structure. It only requires that the client store a location mapping of data blocks. As such, its bandwidth efficiency can be reduced to logarithmic or poly-logarithmic if the client wishes to recursively store the location mappings as another oblivious tree. This construction has the best bandwidth efficiency among the known ORAM protocols and requires little storage on the client's part. While it is very good in the secure processor setting, however, it still suffers from a complex memory block eviction strategy ($O(D \log ^{2} N)$) that, while more optimal for certain problems, does not work well in the multi-party computation setting.

Circuit ORAM~\cite{circuit-oram} is a related ORAM variant which achieves a near-optimal circuit size for realistic memory block sizes. Also tree-based, it takes advantage of preemptive metadata scans to complete the eviction algorithm in one pass, substantially improving the eviction efficiency. This makes it more useful for multi-party computation between servers than Path ORAM.

\section{Two-Party Computation}

Two-Party Computation (2PC) protocols provide a way for two parties $A$ band $B$ to compute a mutual result $f(a,b)$ without revealing their respective secret inputs $a$ and $b$ to each other~\cite{twopc}. 

Both parties start by agreeing on the evaluating function $f$ and obfuscating their secret inputs, most commonly by XOR-ing with a random value. $A$ generates garbled circuits~\cite{garble} that compute $f$ (as well as the required commitments for consistency checks), while $B$ evaluates and decodes the result based on the information sent by $A$.  

Privacy is preserved because each party is involved in a different share of the computation. $A$ knows the random values used in generating the garbled circuit but never comes in contact with input $b$. $B$ receives $A$'s inputs, but does not know what it is because $A$ is in charge of the scrambling and encryption.

\section{Oblix}

Oblix~\cite{oblix}, or OBLivious IndeX, is a space- and time- efficient search index that neither leaks access patterns nor result size. It works by maintaining an obliviously sorted multimap, which essentially links keys to values while ensuring that any given client requests are computationally indistinguishable from random requests. To do so it requires many of the same cryptographic techniques that we will use in our proposed solution, including garbling, ORAM, Merkle hash trees, and oblivious data structures (including the obliviously sorted multimap). 

Oblix has been demonstrated to be useful in private retrieval of public keys in the case of certificate transparency, and in fact solves the basic problem statement presented at the beginning of this paper.  However, it does rely on trusting secure hardware enclaves, which our solution will avoid. 

%% file: src/design.tex
\chapter{Design}
\label{sec:design}

\section{Security}

\subsection{Threat Model}

We assume that we have access to two \emph{non-colluding} servers. These servers can be passive eavesdroppers (honest-but-curious) --- they can observe any local state not secret-shared between the servers and have full access to client communications with itself. 

While requiring two non-colluding servers is perhaps unrealistic in many domains, it fits the certificate transparency system well, where many different companies are providing CT services. These different entities could very well be financial competitors (e.g. Google versus Symantec), and sharing collected datasets would eliminate their commercial advantages. Under our threat model, a CT provider running in this way does not have to be trusted by any clients making requests for certificate proofs.  

We assume that CT security properties are still valid, namely, that unauthorized and/or malicious modifications to certificates will be detected with extremely high probability by a sufficient number of auditing nodes. Thus, we focus specifically on the privacy implications of CT queries.

\subsection{Security Definition}

We define security as preserving the user's privacy. When a user queries the CT server, they should be able to receive the correct proof for the desired domain without revealing to the server what they asked for. 

Even if given two potential domains \textit{a} and \textit{b}, and observing many client requests for one of the domains, a malicious CT log provider should not be able to infer the identity of the queried site any better than a random guesser. Similarly, patterns of requests for the same domain should not reveal any information to the server (e.g.\ a new movie trailer drives a sudden traffic spike to streaming website \texttt{a.com} so any new repeated requests to a given domain is more likely to be a request for that website).  

Our security definition only concerns itself with user privacy, and not that of the server. Since certificate information is meant to be public anyway, it is not a major security threat for a client to know the public keys of domains that they did not ask for. However, we still wish to avoid this scenario to minimize overhead and unnecessary transfers.

\section{Design Proposal}

\subsection{Goals}

We focus on two primary design goals:
\begin{itemize}
    \item Client queries for a domain's certificate do not leak the requested domain to the CT server, as discussed above, without relying on a trusted third party or hardware enclave.
    \item Preserving client privacy imposes reasonable overheads on CT log administration and user requests.
\end{itemize}

\subsection{Oblivious Merkle Trees}

Merkle trees are a special type of binary tree in which a node's value is equal to a hash of its children. Thus, if the value of a node changes even a tiny bit, that change will propagate up to the value of the Merkle tree root~\cite{merkle}.

In the context of certificate transparency, Merkle trees provide a mechanism for users to check the validity of a certificate from a log server with logarithmic efficiency~\cite{ct}. An attacker trying to falsify a certificate cannot realistically find hash collisions well enough to give a valid proof. Merkle trees make it infeasible for an attacker to edit or append certificates without being quickly caught.

However, traditional CT allows the server storing the Merkle tree can easily determine which node a client is requesting the validity proof for (and thus which website a client is trying to visit). This is because the proof consists of the siblings of the nodes on the path from the root to the node the client wishes to validate.

To make certificate transparency work privately, we need to store Merkle trees obliviously. While Oblix~\cite{oblix} is successful at this, our goal is to do the same without relying on secure hardware enclaves. Specifically, we wish to access specific nodes of the Merkle tree, just as in the non-oblivious certificate transparency public key lookup and proof, without revealing which node we accessed.

We do this by using two non-colluding servers that are capable of 2PC. Similar to the Oblix Merkle tree implementation, nodes are identified using an oblivious index~\cite{oblix}. Instead of relying on secure enclave, however, we take advantage of the "separation of powers" provided by two-party computation. 

We store the Merkle tree nodes and certificate data using ORAM~\cite{circuit-oram} on one server, and the position mappings on the other. One server knows the values of the nodes, and the other server knows the positions of the nodes, but neither knows both. Only the user gets both the position and value shares, and can assemble this information privately to generate the Merkle tree proof. This setup still uses the original tree structure of the data nodes and operates at $O(\log n)$ per request.

\begin{figure}
  \includegraphics[width=\linewidth]{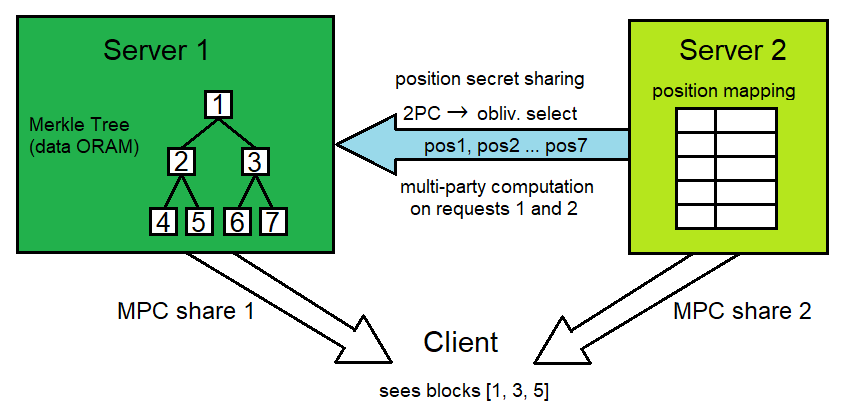}
  \caption{The two servers use two-party computation to create the client's request, without knowing the other's share.}
  \label{fig:merkle}
\end{figure}

\subsection{Scalability}

The crux of our solution to providing private queries on public CT data relies on a scalable private storage mechanism. Many existing private storage solutions (as discussed in Section~\ref{sec:relwork} are inefficient, do not scale to many users, rely on a trusted proxy or specialized hardware, or some combination of these.

Using ORAM and 2PC alone would fulfill the first requirement. However, while oblivious file sharing performs significantly better than existing private file storage systems, its initial version is does not scale to serve the entire world's CT requests. The basic system can only process a single request at a time in serial due to the ORAM construction, so multiple users making many simultaneous requests results in fast-growing latencies. Currently, reading a 4KB file from this basic system takes about 2.5s when there are many files in the system under our testing setup, which we will discuss further in Section~\ref{sec:eval}. 

One key design consideration for CT is that because certificate validation is performed online and in batches, providing latencies that are good enough for real-time browsing is unnecessary in this context.
However, to realistically serve many clients, our system must be able to more gracefully handle multiple requests in parallel.

A primary focus of our project is to modify this existing system to make it more efficiently serve multiple concurrent requests. We do so by precomputing encryption exponentiations and garbled circuits, pipelining relevant stages, and batching similar requests.

%% file: src/implementation.tex
\chapter{Implementation}
\label{sec:implementation}

\section{Implementing Oblivious Merkle Trees}

We make the same assumptions as was mentioned in the threat model in Section~\ref{sec:design}: the client has access to two mutually distrustful servers. These servers can both independently be passively malicious, but they do not collude. In implementing the oblivious Merkle tree, we use C and its associated libraries like Obliv-C~\cite{obliv-c} for the baseline system implementation. 

One server is responsible for storing the actual data, that is, the certificate transparency Merkle tree nodes. The other server knows the random position mapping for the nodes (analogous to the oblivious index of Oblix). By using oblivious transfer, the two servers can secret share the position mapping and use two-party computation to generate the two shares of the final value. As the client is the only one that possesses both shares simultaneously, they are they only one who receives the final result. A basic version of this protocol is shown in Algorithm~\ref{twopc}. 

Once a node has been accessed, it must be evicted and and randomly placed back in the server as in Circuit ORAM to prevent the server from linking a previous request to a later request. Otherwise, a server will be able to tell that the user has accessed the same node multiple times. The eviction protocol is shown in Algorithm~\ref{ORAM_eviction}. \\
\\

\begin{algorithm}
\SetAlgoLined
\label{twopc}
\textbf{Input:} server1 (Merkle nodes), server2 (position mappings) \\
\ \\
circuit = server1 and server 2 agree on a circuit to compute the Merkle proof\\
garbledCircuit, labels = server2.garble(circuit) \\
encryptedPosmap = server2.encrypt(positionMap, labels) \\
encryptedMerkle = server1.obliviousRequest(ctMerkle, labels) \\
\ \\
server2.send(server1, garbledCircuit) \\
server2.send(server1, encryptedPosmap) \\
encryptedProof = server1.eval(garbledCircuit, encryptedMerkle, encryptedPosmap) \\
outputLabel = server1.decrypt(garbledCircuit, encryptedProof) \\
\ \\
share1 = server1.format2PC(outputLabel) \\
server1.send(client, share1) \\
share2 = server2.format2PC(garbledCircuit) \\
server2.send(client, share2) \\
merkleProof = client.recompute(share1, share2) \\
\ \\
\caption{The basic strategy for requesting the Merkle tree proof in our oblivious CT scheme, as seen in 2PC~\cite{twopc} and oblivious transfer~\cite{obliv-transfer}. For simplicity we exclude the standard permissions and correctness checks.}
\KwResult{Obliviously fetches and returns the required blocks for a Merkle tree proof.}
 
\end{algorithm}
\

\begin{algorithm}
\SetAlgoLined
\label{ORAM_eviction}
\textbf{Input:} CertificateNode \\
\ \\
 merkTree = Server1.initMerkleTree()\;
 posMap = Server2.initPositionMapping()\;
 \textit{List} proofNodes = [required nodes for Merkle proof of CertificateNode] \\
 \ \\
 \For{node in proofNodes}{
    nodePosition = posMap.getPosition(node) \\
    path = path to nodePosition \\
    deepest = \textit{Prepare deepest legal node for eviction} \\
    target = \textit{Prepare target block for writing}
    \ \\
    holdBlock = $\perp$ \\
    destination = $\perp$ \\
    \For{i in [0 ... length of path]}{
        writeBlock = $\perp$ \\
        \If{holdBlock != $\perp$ and i == destination}{
            writeBlock = holdBlock \\
            holdBlock = $\perp$ \\
            destination = $\perp$ \\
        }
        \If{target != $\perp$}{
            holdBlock = \textit{read and remove deepest legal block in} path[i] \\
            destination = target[i] \\
        }
        \If{writeBlock != $\perp$}{
            path[i].write(writeBlock) \\
        }
    }
 }
 \ \\
 \caption{\texttt{evict()}: The basic eviction strategy when obliviously requesting the Merkle tree proof in our private CT scheme, as seen in 2PC~\cite{twopc} and Circuit ORAM~\cite{circuit-oram}.}
 \KwResult{After fetching the required blocks for the Merkle tree proof, reshuffles the blocks in ORAM.}
 
\end{algorithm}

\pagebreak
\section{Implementing Scaling}

In this section, we describe our changes to the baseline oblivious file sharing system implementation that allow it to scale more efficiently when servicing multiple requests concurrently. 

\subsection{Precomputation}

Oblivious transfer protocols use asymmetric encryption and garbled circuits. Both of these processes are computationally expensive and used many times in the sharing protocol, but are also independent of the inputs and CT itself. Thus, they can be precomputed offline, separate from the actual requests. This should improve both individual request latency as well as the total throughput of the system.

Modern asymmetric encryption schemes involve computing exponents of very large numbers. This can be done in a separate thread from the one that services requests. Afterwards, the values can be retrieved as needed. We only need to be careful to protect the precomputed exponentiations from cache timing attacks.

Garbled circuits~\cite{garble} are also very time-consuming because they cannot be reused, and a new one must be computed for each transfer. Having one server generate a garbled circuit and then transfer it to the other only when needed introduces significant latency and could be offloaded to another set of servers that computes them offline or in the background. Again, this can be computed simultaneously with and separately from the rest of the transfer protocol. While the garbled circuit itself is important for the transfer, the composition of the circuit itself is independent of the request so long as it takes the proper number of inputs, is properly randomized, and changed for every request. We generate the garbled circuits in a separate thread on demand.

\subsection{Pipelining}

The system's server-side protocol to access files can be seen as a series of sequential stages. Some of these stages must be performed sequentially for a single request, while others can happen in parallel.

The nine main stages of a single request are listed below:

\begin{enumerate}
    \itemsep0em 
    \item Receiving client requests and validating commitments.
    \item Performing permissions checks on the user and files.
    \item Obliviously fetching the CT Merkle tree node positions.
    \item Obliviously fetching the CT Merkle tree proof nodes.
    \item Generating a permutation to shuffle the paths in ORAM.
    \item Applying the generated permutation to ORAM.
    \item Computing the 2PC result shares for each server.
    \item Secret sharing the result to the client.
    \item Recomputing the result based on the given secret shares.
\end{enumerate}

The system cannot concurrently fetch a new path from ORAM while the previous permutation has not been applied to ORAM. This results either in a loss of obliviousness if the permutation is not performed at all, or an incorrect lookup if the two operations happen concurrently. 

On the other hand, stages 1-2 are completely independent of the following stages, as long as we are careful not to finish the revealing process before the validations and permissions checks are finished. This suggests that the system could benefit from pipelining these stages, or overlapping the execution stages across requests to provide partial parallelism and improve performance when servicing multiple requests. 

Our design is limited by the fact that Obliv-C~\cite{obliv-c} currently only supports a single thread running two-party computations. This means, for example, that while stages 2 and 3 could be pipelined, we are not able to do so as we must perform all 2PC in a single thread. Adding parallelism would require major modifications to Obliv-C that are outside the scope of this project.

Our implementation consists of three logical groups of stages that run in separate threads: stage 1, stages 2-6, and stages 7-9. While performing all 2PC in a single stage is not ideal, in fact the group consisting of stages 2-6 is the bottleneck, so splitting stages 2-6 would not result in a significant speedup.

\subsection{Batching}

In scenarios such as Certificate Transparency, client latency is a small concern compared to the impact of raw throughput on scalability. We explore how to widen bottlenecks in the pipeline discussed above by batching and processing multiple independent requests in parallel. While this approach results in longer per-request processing times, we can increase the ability of server pairs to handle higher load.

We first enable the system to successfully process interleaved requests -- two or more simultaneously transmitted query shares may arrive at a server in any order and still be processed correctly. We synchronize and update the pending, incomplete queries by their commitment values and expose a queue-based interface to the main file access handler, similar to the queues used to implement pipelining. The handler withdraws requests from the queue, synchronizes with the second server, and processes the query as normal. This capability allows for processing sets of batched requests, simply by withdrawing more than 1 query at a time from the incoming queue.

Interacting with the ORAM requires a set of individual path queries, so it is difficult to fully merge a batch of queries. However, we observe that the Circuit ORAM's tree-based structure results in notable overlap between different operations. In addition, while computing over the small file-index ORAM is quite quick in a 2-party computation setting, operations on the larger filesystem tree is more costly. Rather than synchronizing the data ORAM serially for every request, we compose the changes caused by the batch as a whole to require at most one update per block in the tree.

As a simple example, imagine a query that results in a block eviction into the root, followed by a second query whose result moves the same block from the tree root to a bucket at a leaf. Operating sequentially, we would follow these exact steps and relocate the same data block twice. On the other hand, applying multiple operations to the index ORAM tree at once, while performing the appropriate flushes, can yield more efficient updates, in this case directly from the stash into a leaf bucket.

Implementing this requires a fundamental change to the type of permutation generated by the 2PC before it is applied by each server to the data ORAM. Whereas previous permutations consist solely of two paths intersecting only at the stash and root, composing the result of many individual evictions requires combining a subtree of varying shape, depending on the particular slice of the eviction schedule. Maintaining a full-tree permutation representation is not memory efficient or necessary; instead, we continue to track changes to the ORAM tree in a global context (rather than relative to the current eviction path), but in a lightweight C macro-based map \cite{uthash} that scales the batch size.

We found that removing path-relative tracking reduces the complexity of index management required to track ORAM changes, as we can reset the index values at any point and derive any changes made to the entire tree after some amount of processing. This structure can be collapsed into an array with explicit index labels in order to create permutation shares for the individual machines and apply the tracked changes to the filesystem itself.

%% file: src/eval.tex
\chapter{Evaluation}
\label{sec:eval}

Our primary evaluation mechanism was observing the behavior of the modified oblivious file sharing system under load from multiple concurrent clients, requesting as many randomly-selected files as possible over a 60 second window, extracting latency and throughput behavior as the number of competing clients (e.g. potential CT-enabled browsers) increases. 

For our testing we used an Ubuntu m5d.xlarge NVMe SSD type AWS instance running on one core with 150GB memory. Without any improvements at all, a single serial request may take 2.5s or more on our experimental setup depending on the request size and database size. 

While we did observe positive trends associated with our systems-level optimizations, they did not significantly change the overall complexity of processing each request. Due to general resource constraints and the aforementioned request complexity, we had to limit our evaluation to small numbers of parallel clients instead of the larger scale that would be more realistically seen in a potential real-world datacenter.

\section{Precomputation}

As discussed previously, encryption caching by itself provides a little bit of speedup but does not change the overall complexity. Similarly, precomputing the garbled circuits technically does not change the big-O complexity, but in practice shaves off more time than caching of exponentiation results. Together, they make the latency more bearable.

\section{Pipelining}

\begin{figure}
  \includegraphics[width=\linewidth]{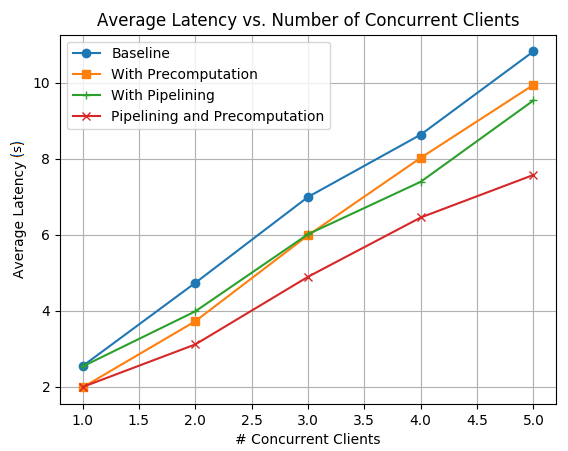}
  \caption{Average client latency under contention, with pipelining enabled and disabled.}
  \label{fig:latency}
\end{figure}

Figure~\ref{fig:latency} evaluates the impact of encryption and garbled circuit precomputation/caching and the pipelining mentioned in Section \ref{sec:implementation} on client latency. The filled in measurements represent clients running a two-stage pipeline, and indicate that such a technique can slightly increase the effectiveness of other optimization. Specifically, as the number of concurrent clients increases, simply pre-computing exponentiations does not benefit client latency, but pipelining lowers average latency by a constant factor and benefits exponentiation. This is likely due to the fact that the caching in the baseline serial implementation only occurs when there are no requests to be handled, whereas the inclusion of multiple pipeline stage threads allows exponentiation to persist in the background. Thus, we believe the ability to concurrently schedule different tasks can directly impact overall system effectiveness.

\begin{figure}
  \includegraphics[width=\linewidth]{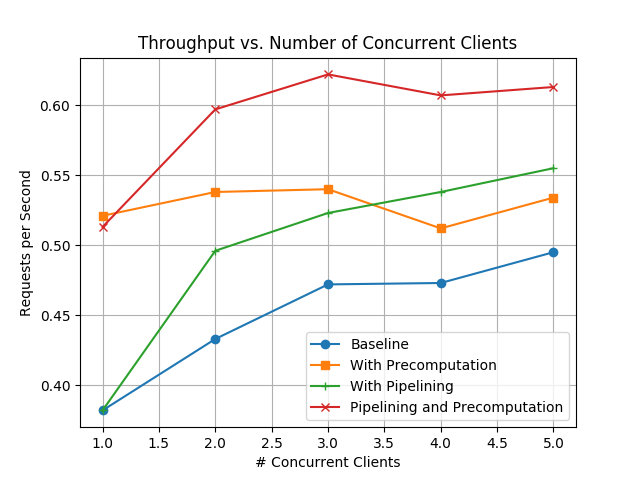}
  \caption{Server throughput as the number of parallel requests grows, with pipelining enabled and disabled.}
  \label{fig:tput}
\end{figure}

Figure~\ref{fig:tput} tracks server-side throughput with multiple outstanding requests (in these experiments, the clients only send one query at a time). As expected, pipelining performs at the same level as the baseline when only one request is processed at a time since each of the other stages are empty. Throughput rises drastically with two parallel requests (filling up both current pipeline stages) and flattens out with increased load as the pipeline is completely full. As mentioned earlier, a more flexible MPC implementation with multithreading capabilities would enable much finer granularity (more pipeline stages) and therefore more significant performance gains. The best-performing combination in terms of throughput was again both precomputation and pipelining with an approximate 22\% performance gain, although a pure pipelining-based approach was eventually able to exceed the initial raw performance boost offered by pre-computation. Again, as with latency, which these improvements do make the system faster, they do not perform significantly better as more and more requests are provided.

\section{Batching}

We successfully modified the oblivious Circuit ORAM implementation used by our file sharing system to track and collapse subtree-based modifications for arbitrary batch sizes as described in Section \ref{sec:implementation}. However, we encountered an late implementation issue using Obliv-C to expose the per-server permutations to each party in the 2PC. Thus, while we can demonstrate that such a method will work in practice, we were not able to perform the same evaluation as above while batching was enabled.

%% file: src/conclusion.tex
\chapter{Conclusion and Open Questions}
\label{sec:conclusion}

We have shown that while conducting private queries on public databases is possible without the use of hardware enclaves, the opportunity for systems-based approaches to make scalability gains is significant. Neither our upgraded oblivious file sharing system nor similar privacy-preserving systems are ready to serve hundreds of millions of users, but they provide an important first step. This demonstrates that improvement of performance while preserving privacy is possible, and more work will make our goal a reality.

Currently, it is possible that specific individuals with the right personal incentives, such as human rights workers or asylum seekers, would be willing to deal with the extra latency for the sake of privacy and security.

To serve users at scale, our system needs to be able to scale to multiple machines. There has been work on distributed ORAM in the past such as such as ObliviStore~\cite{oblivistore} and TaoStore~\cite{tao}, but designing a similar construction that works with 2PC and in the larger oblivious file sharing system presents a significant challenge.

One possible area for even future improvement is making the requests for the Merkle tree proof more efficient. Currently, we individually request each node in the Merkle tree proof. However, being able to do the requests and evictions in one pass would reduce the overall request time. We must be careful, however, to ensure that none of the requests reveal anything about the others, and thus hide from the server the path of the Merkle tree proof and the ultimate value of the originally requested certificate at the leaf. 

As a side note, it would be helpful to see how our implementation holds up against a real-world CT server. The publicly available source code on Github by Google has two versions, a deprecated C version and a newer Golang version. Because the existing oblivious file sharing system and most of our codebase has already been implemented in C, we have been working with the deprecated C version of CT as well for simplicity's sake.

While the protocols often used in such projects are highly studied and constructed for efficiency, real-world cryptosystems are only broadly adopted if they are practical to deploy and operate at reasonable cost. We hope to continue our work on both privacy-sensitive, secure internet infrastructure, including Certificate Transparency, and scalable private systems in general.

%% file: src/acknowledgements.tex
\chapter{Acknowledgements}
\label{sec:relwork}

I would like to thank Jean-Luc Watson and Edward Oakes, for their great contributions to the system-level improvements discussed in this paper.

I would also like to thank my advisors, Doug Tygar and Raluca Ada Popa, for their support and advice throughout this project.